\documentclass[twocolumn]{aastex631}

\usepackage{xcolor}

\usepackage{booktabs}
\usepackage{multirow}
\usepackage{array}

\shorttitle{AASTeX v6.3.1 Sample article}
\shortauthors{Wang et al.}

\graphicspath{{./}{figures/}}

\begin{document}

\title{HST Confirms Sub-5 kpc Dual Quasar Pairs at Cosmic Noon}

\author[0009-0000-4339-8117]{Qian Wang}
\affiliation{School of Physics and Technology, Wuhan University, Wuhan 430072, China}

\author[0000-0001-8917-2148]{Xuheng Ding}\thanks{dingxh@whu.edu.cn}
\affiliation{School of Physics and Technology, Wuhan University, Wuhan 430072, China}
\affiliation{Kavli Institute for the Physics and Mathematics of the Universe (Kavli IPMU, WPI), UTIAS, Tokyo Institutes for Advanced Study, University of Tokyo, Chiba, 277-8583, Japan}

\author[0000-0002-0000-6977]{John Silverman}
\affiliation{Kavli Institute for the Physics and Mathematics of the Universe (Kavli IPMU, WPI), UTIAS, Tokyo Institutes for Advanced Study, University of Tokyo, Chiba, 277-8583, Japan}
\affiliation{Department of Astronomy, School of Science, The University of Tokyo, 7-3-1 Hongo, Bunkyo, Tokyo 113-0033, Japan}
\affiliation{Center for Data-Driven Discovery, Kavli IPMU (WPI), UTIAS, The University of Tokyo, Kashiwa, Chiba 277-8583, Japan}
\affiliation{Center for Astrophysical Sciences, Department of Physics \& Astronomy, Johns Hopkins University, Baltimore, MD 21218, USA}

\author[0000-0002-7738-6875]{J. Xavier Prochaska}
\affiliation{Department of Astronomy and Astrophysics, University of California, Santa Cruz, 1156 High Street, Santa Cruz, CA 95064, USA}
\affiliation{Kavli Institute for the Physics and Mathematics of the Universe (Kavli IPMU, WPI), UTIAS, Tokyo Institutes for Advanced Study, University of Tokyo, Chiba, 277-8583, Japan}
\affiliation{Division of Science, National Astronomical Observatory of Japan, 2-21-1 Osawa, Mitaka, Tokyo 181-8588, Japan}

\author[0000-0002-8460-0390]{Tommaso Treu}
\affiliation{Department of Physics and Astronomy, University of California, Los Angeles, 430 Portola Plaza, Los Angeles, CA 90095, USA}

\author[0000-0002-4176-9145]{Hassen M. Yesuf}
\affiliation{Key Laboratory for Research in Galaxies and Cosmology, Shanghai Astronomical Observatory, Chinese Academy of Sciences, 80 Nandan Road, Shanghai 200030, China}

\author[0000-0003-4700-663X]{Andrew D. Goulding}
\affiliation{Department of Astrophysical Sciences, Princeton University, 4 Ivy Lane, Princeton, NJ 08544, USA}

\author[0000-0001-6186-8792]{Masatoshi Imanishi}
\affiliation{National Astronomical Observatory of Japan, National Institutes of Natural Sciences (NINS), 2-21-1 Osawa, Mitaka, Tokyo 181-8588, Japan.}

\author[0000-0003-3954-4219]{Nobunari Kashikawa}
\affiliation{Department of Astronomy, Graduate School of Science, The University of Tokyo, 7-3-1 Hongo, Bunkyo-ku, Tokyo 113-0033, Japan}
\affiliation{Research Center for the Early Universe, The University of Tokyo, 7-3-1 Hongo, Bunkyo-ku, Tokyo, 113-0033, Japan}

\author[0000-0002-9267-2677]{Issha Kayo}
\affiliation{Department of Liberal Arts, Tokyo University of Technology, Ota-ku, Tokyo 144-8650, Japan}

\author[0000-0002-4052-2394]{Kotaro Kohno}
\affiliation{Institute of Astronomy, Graduate School of Science, The University of Tokyo, 2-21-1 Osawa, Mitaka, Tokyo, 181-0015 Japan}
\affiliation{Research Center for the Early Universe, Graduate School of Science, The University of Tokyo, 7-3-1 Hongo, Bunkyo-ku, Tokyo 113-0033, Japan}

\author[0000-0002-4359-5994]{Kai Liao}
\affiliation{School of Physics and Technology, Wuhan University, Wuhan 430072, China}

\author[0000-0001-5063-0340]{Yoshiki Matsuoka}
\affiliation{Research Center for Space and Cosmic Evolution, Ehime University, 2-5 Bunkyo-cho, Matsuyama, Ehime 790-8577, Japan}

\author[0000-0002-0106-7755]{Michael A. Strauss}
\affiliation{Department of Astrophysical Sciences, Princeton University, 4 Ivy Lane, Princeton, NJ 08544, USA}

\author[0000-0002-2185-5679]{Shenli Tang}
\affiliation{University of Southampton, University Road, Southampton, SO17 1BJ, United Kingdom}

\begin{abstract}
During cosmic noon ($z\sim1-3$), when both star formation and black hole growth peaked, galaxy mergers are predicted to trigger dual active galactic nuclei (AGN) that eventually coalesce as supermassive black hole (SMBH) binaries. However, observations of dual quasars with sub-5 kpc separations—the critical phase preceding final coalescence—have remained challenging due to angular resolution limitations. We present the discovery and confirmation of two sub-arcsecond dual quasars at $z>1$, selected from 59,025 SDSS quasars, which fall within the footprint of the Hyper Suprime-Cam Survey. Using high-resolution \textit{Hubble Space Telescope} (HST) imaging and slitless spectroscopy, we confirmed SDSS~J1625+4309 ($z=1.647$, separation 0\farcs55/4.7 kpc) and SDSS~J0229$-$0514 ($z=3.174$, separation 0\farcs42/3.2 kpc), probing the sub-5 kpc separation regime. Through novel combination of WFC3/IR direct imaging (F140W) and grism spectroscopy (G141), we resolve both components morphologically and spectroscopically confirm their dual nature via detection of H$\beta$+[O\,\textsc{iii}] and Mg\,\textsc{ii} emission lines in each nucleus. Two-dimensional image decomposition reveals distinct host galaxy morphologies: J1625+4309 shows an extended, disturbed structure ($R_e$=4.7 kpc) indicative of an ongoing major merger, while J0229$-$0514 exhibits a compact host ($R_e$=1.4 kpc) suggesting an advanced coalescence stage. Black hole mass estimates based on virial relations yield $\mathcal{M}_{\mathrm{BH}} \sim 10^{8.1}-10^{8.7} M_\odot$ with line-of-sight velocity offsets of $(0.7\pm0.1)\times10^{3}$ km s$^{-1}$ and $(1.0\pm0.2)\times10^{3}$ km s$^{-1}$, respectively. These confirmations directly constrain the frequency and properties of close dual quasars, opening new avenues for studying SMBH mergers at cosmic noon.
\end{abstract}

\keywords{Double quasars (406); Quasars (1319); Galaxy mergers (608)}

\section{Introduction} \label{sec:intro}

Galactic mergers serve as a fundamental driver of galaxy evolution in the universe. During this process, the two supermassive black holes (SMBHs) at the centers of the host galaxies approach each other, potentially forming a dual SMBH system or ultimately merging \citep{Rosas-Guevara_2019, Saeedzadeh_2024, Vazquez_2025}. When these SMBHs simultaneously exhibit active accretion within separations ranging from kiloparsecs (kpc) to parsecs (pc), they form dual active galactic nuclei or quasars (dual AGN/quasar), which signify the late stages of galaxy mergers and herald the eventual coalescence of the SMBHs \citep{Comerford_2015, Li_2025, Rosa_2019}. The detection and study of dual AGN/quasar systems provide critical insights into: (i) the triggering mechanisms of SMBH accretion \citep{Capelo_2015, Capelo_2017, Ruby_2024}, (ii) the co-evolution of black holes and their host galaxies \citep{Volonteri_2016, PuertoSanchez_2024}, and (iii) the physical processes surrounding SMBH mergers, such as accretion disk interactions and gravitational wave emission \citep{Sandoval_2024}, thus enabling predictions of low-frequency gravitational wave event rates and offering key clues to galaxy formation and evolution in the early universe \citep{Wang_2025, Burross_2025, Dodd_2025}. Cosmological models further predict a substantial population of such dual SMBHs in shared host galaxies, which will ultimately merge and generate intense gravitational waves \citep{Mannucci_2022}.

The observational identification of dual quasars faces significant challenges due to their small angular separations and potential confusion with gravitationally lensed systems \citep{tang_2021} and projected quasar pairs that are not physically bound \citep{Anguita_2018}. Current detection methods primarily employ several complementary approaches: (i) high-resolution imaging, (ii) double-peaked emission line spectroscopy, (iii) very long baseline interferometry (VLBI), (iv) Gaia astrometry, and (v) JWST integral field spectroscopy. Each technique has demonstrated success, but also carries specific limitations. High-resolution imaging with facilities like the \textit{Hubble Space Telescope} (HST) remains the gold standard for resolving close-separation dual quasars, having identified systems with projected separations as small as 0\farcs13 ($\approx$430 pc at z=0.2) \citep{goulding_2019}. Spectroscopic methods relying on double-peaked emission lines [O\,\textsc{iii}] $\lambda$ 5007 \citep{Liu_2010, Frey_2012, Zheng_2025} can suggest dual nuclei but require confirmation due to potential contamination from jet-driven outflows or rotating disks \citep{Gabanyi_2016, Liu_2018, Shen_2011_b}. Although VLBI achieves unparalleled angular resolution in the radio bands, its utility is limited by potential discrepancies with optical data \citep{Veres_2021,Koss_2015}. Gaia astrometry provides statistical samples of wide-separation pairs \citep{KroneMartins_2019,Hwang_2020,Chen_2023_11,Gross_2025,Jiang_2025}, and JWST enables revolutionary studies of high-$z$ dual systems through infrared spectroscopy \citep{Perna_2023}.

Cosmic noon ($z \sim 1-3$) is a particularly important epoch, when cosmic star formation and SMBH growth peak \citep{Schreiber_2020, Mezcua_2024}, and the detection of close-separation ($<5$ kpc) dual quasars provides crucial insights into galaxy-SMBH co-evolution \citep{Ishikawa_2024, Mannucci_2022}. While extensive surveys like HSC-SSP have identified hundreds of wide-separation ($>10$ kpc) candidates \citep{Silverman_2020} and JWST has revealed obscured systems through infrared spectroscopy \citep{Shen_2021, Ishikawa_2024, Chen_2024}, the sub-5 kpc regime at $z>1$ remains largely unexplored due to angular resolution limitations. Theoretical models predict peak AGN activity during close encounters ($3-5$ kpc) preceding SMBH coalescence \citep{Hopkins_2018, HutchinsonSmith_2018, Volonteri_2022}, but the current observations are still far from sufficient.

In this work, we overcome the observational limitations through a HST program which applies a novel combination of direct WFC3/IR imaging with 0\farcs06 resolution (F105W/F140W) and slitless grism spectroscopy (G102/G141), enabling morphological separation of pairs up to 0\farcs45 (2.5 kpc at $z=2$) and providing simultaneous spectral confirmation of both components for lens discrimination through matched broad-line profiles. We represent a systematic search for sub-6~kpc dual quasars at $z>1$, filling a critical parameter space for understanding the binary evolution of SMBHs.

We describe the sample selection and data observation in Section \ref{sec:sample}. The detailed data reduction process is provided in Section \ref{sec:reduc}. The nature of the two systems is discussed in Section \ref{sec:confirm}, based on their imaging and spectroscopic properties. A summary appears in Section \ref{sec:summary}. We adopt the cosmological parameters \( H_0 = 70~\mathrm{km\,s^{-1}\,Mpc^{-1}} \), \( \Omega_{\mathrm{M}} = 0.3 \), and \( \Omega_{\Lambda} = 0.7 \).

\section{Sample selection and HST observation} \label{sec:sample}

\begin{figure*}[htbp!]
    \centering
    \includegraphics[width=0.9\textwidth]{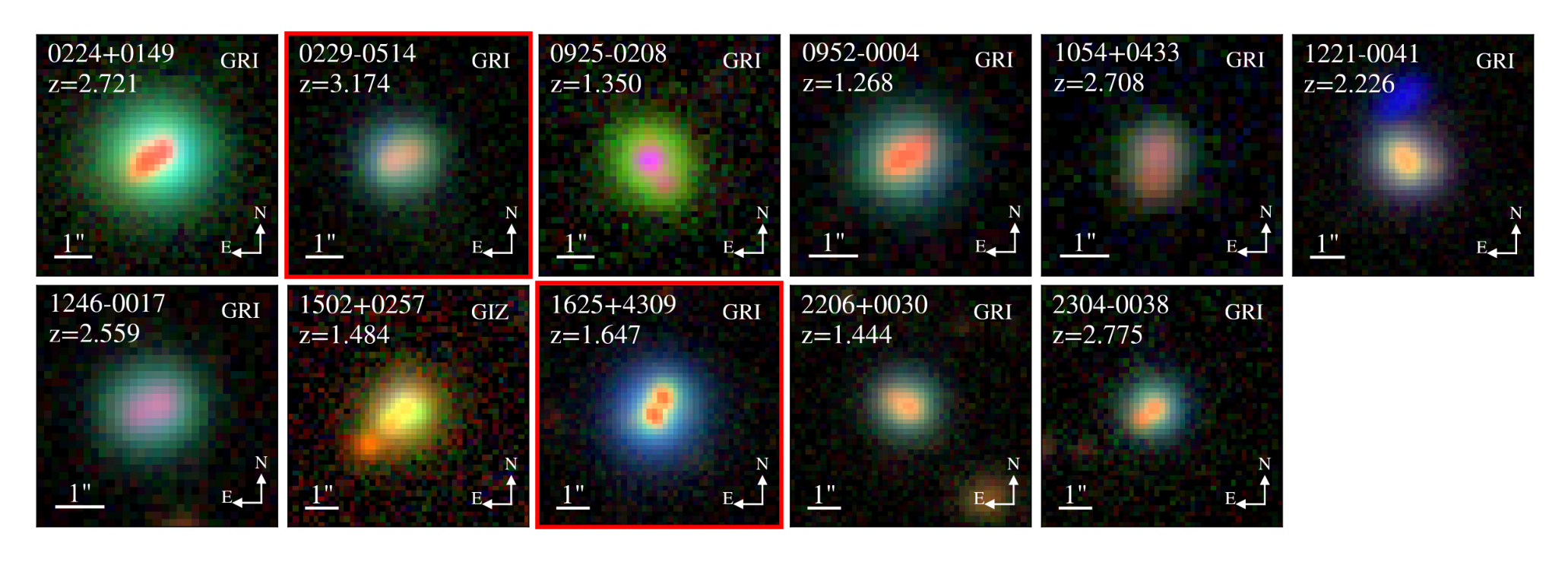} 
    \caption{The pseudo-color HSC images for all 11 candidates and redshift information with their names, redshifts, and combined HSC bands indicated. The two confirmed dual quasar systems are marked with red boxes (i.e., SDSS~J0229$-$0514 at $z=3.174$ and SDSS~J1625$+$4309 at $z=1.647$).}
    \label{fig:HSC_color_images}
\end{figure*}

The sample for this study was constructed from a targeted search for dual quasar candidates within 500 square degrees of Hyper Suprime-Cam (HSC) imaging \citep{aihara_hyper_2018}, focusing on Sloan Digital Sky Survey (SDSS) quasars \citep{shen_2011_a}. % at redshift $z>1$ [ref]. 

In total, 59,025 unsaturated SDSS quasars are located within the HSC footprint. Based on an automated 2D image fitting and detection algorithm optimized for the HSC point spread function (PSF), we initially identified 386 SDSS quasars exhibiting two bright sources or with spatially distinct features as reported in \cite{Silverman_2020, tang_2021}. We flagged them as potential dual quasar candidates. Each candidate was then subjected to a thorough visual inspection to exclude sources with stellar colors ($g-r>1$), extended morphological features indicative of lower-redshift galaxies, or companions fainter than an i-band magnitude of 23.5. This magnitude cutoff was applied to ensure that the companion object is sufficiently bright for spectroscopic follow-up with HST, as fainter sources would not yield adequate signal-to-noise ratio (SNR) for the detection of broad emission lines. 
To select the physically associated dual quasars with the closest separation, we further restricted our selection to 25 candidates with projected separations less than 0\farcs7, corresponding to physical scales of less than 6 kpc.

We then considered the scheduling constraints, orbital visibility, and the need to prioritize systems with the highest likelihood of being physically associated dual quasars, which yielded a final sample of 11 high-priority candidates (with SDSS-measured redshifts ranging from 1.2 to 3.2, as shown in Figure \ref{fig:HSC_color_images}) for HST follow-up. As a final check, we cross-matched these 11 candidates with the GAIA DR3 catalog and confirmed that none of the dual components had a significant parallax measurement that would identify them as foreground stars. We acknowledge that Gaia parallax detections are not feasible for all stars, and this final check was instead intended as a precautionary step.

Observations were conducted with HST/WFC3 in slitless grism mode (Program ID GO-17143, PI: Xuheng Ding), utilizing either the G102 ($0.8-1.15\mu$m) or G141 ($1.1-1.7\mu$m) grisms, selected to match the redshifted positions of key diagnostic emission lines for robust AGN identification and redshift confirmation. Four targets at $z\sim2.5-3.1$ were observed with G102 to capture Mg\,\textsc{ii} $\lambda2800$ emission, while 6(1) systems  at $1.2<z<2.5$($z>3.1$) were observed with G141 to target H$\beta$+[O\,\textsc{iii}] $\lambda5007$(Mg\,\textsc{ii} $\lambda2800$) features. Each observation included complementary direct imaging with F105W or F140W filters to verify source morphologies and rule out lensing galaxies. 
Spectroscopic exposure times were carefully optimized to set as 30 minutes, driven by the need to reach a 10$-\sigma$ S/N for 10/11 candidates, which have both pairs brighter than 22.5 mag. For J1054+0433, which has a fainter pair as 23.3 mag, 1.5 hours exposure is sufficient to reach an 8$-\sigma$ S/N. We estimate exposure times required to meet these goals with either the G102 (4 cases) and G141 (7 cases) grism for all 11 dual quasar candidates. To maximize observational efficiency, we implemented an alternating sequence of short direct imaging exposures (100s) followed by longer grism exposures ($\sim$500s), effectively eliminating buffer dump overheads. This observational strategy, combined with our target selection criteria, required a total of 12 Hubble Space Telescope orbits to complete the program.

\section{Data reduction} \label{sec:reduc}

The HST/WFC3 direct imaging and slitless spectroscopic data were reduced using a standard pipeline. Individual spectroscopic frames were then combined with the \texttt{astrodrizzle} algorithm to produce a single, high SNR image. A matching drizzle procedure was applied to the direct imaging data, aligning the frames to the reference spectroscopic image to ensure spatial consistency. 
After drizzle processing, the spectral data reach a final scale of 46.5 \AA/pixel, while the direct imaging retains the native 0.064\arcsec/pixel resolution.
We employed \texttt{photutils} to perform background subtraction on the drizzled direct images, followed by the generation of deblended segmentation maps to identify and isolate distinct sources within the field.

\begin{figure*}[htbp!]
\centering

\begin{minipage}{0.9\textwidth}
    \centering
    \includegraphics[width=\textwidth]{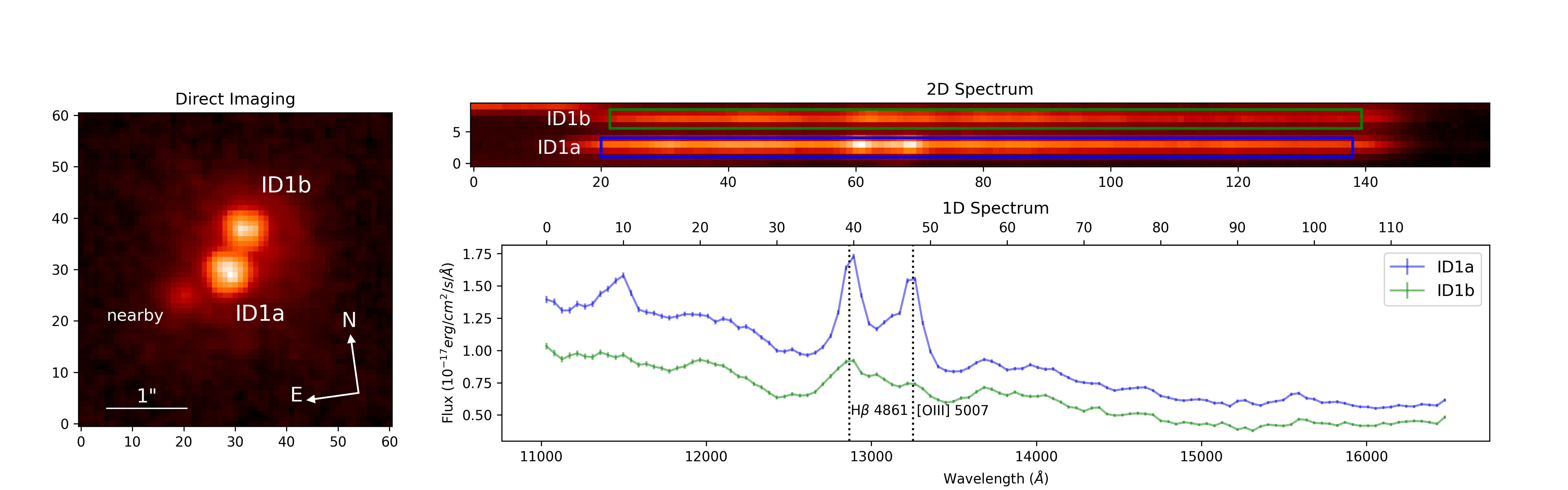}
    \\(a) J1625+4309
\end{minipage}

% \vspace{0.3cm}

\begin{minipage}{0.9\textwidth}
    \centering
    \includegraphics[width=\textwidth]{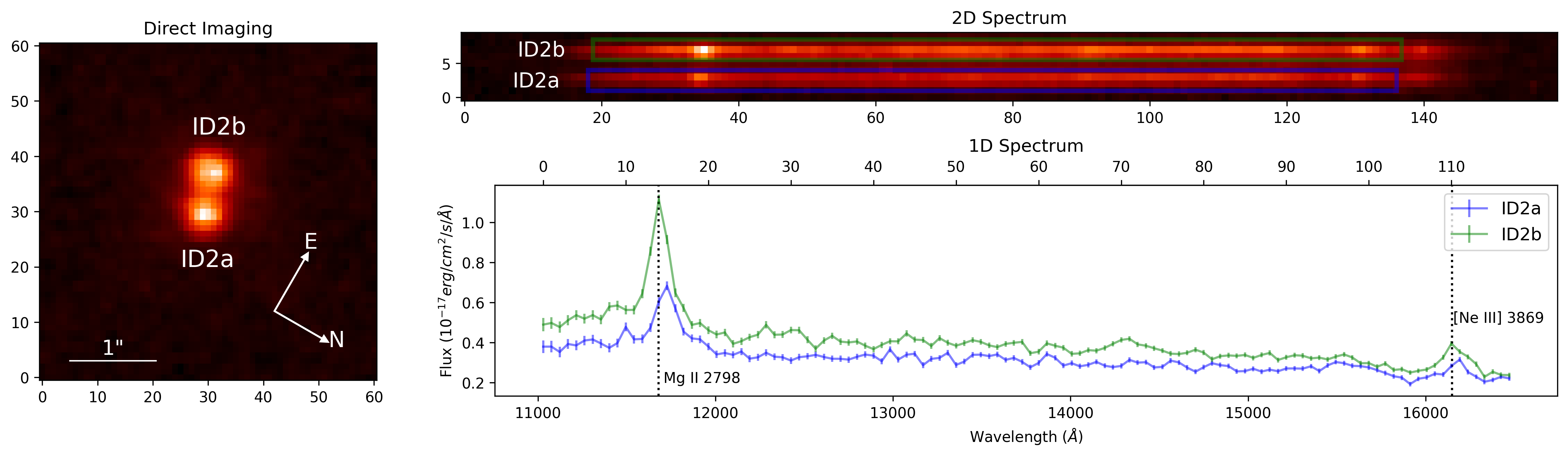}
    \\(b) J0229$-$0514
\end{minipage}

\caption{
HST observations of the two confirmed dual quasar systems. For each system: direct F140W imaging ({\it left}),
two-dimensional spectra ({\it upper right}), and one-dimensional spectra ({\it lower right}). All four components (ID1a, ID1b, ID2a,
ID2b) are clearly resolved.
}
\label{fig:data_reducted}
\end{figure*}

We take advantage of the high-resolution spatial ability of HST to extract spectra of both components of the dual AGN candidates. Using the \texttt{HSTaXe} \citep{aXe} software package, we generated calibrated 2D spectra for all dual quasar candidate systems. %The environment was configured by setting up requisite directories and environment variables, and an association file linking the direct and spectroscopic data was generated. 
%The extraction was conducted using the \texttt{axeprep} and \texttt{axecore} routines, yielding two-dimensional spectra. 
Following verification of the linear response between 2D spectra and their automatically extracted 1D counterparts, we performed optimized spectral extraction for individual quasar components. The extraction was performed at the central 3-pixel region of each component's 2D spectrum (see Figure~\ref{fig:data_reducted}) --- this 3-pixel width was chosen to ensure that it lies beyond the $2\sigma$ line width contamination from the adjacent component.

After extracting the initial 1D spectra, we further performed 2D image decomposition of the drizzled imaging data using \texttt{Galight}~\citep{2020ApJ...888...37D} to quantify the quasar luminosity in each target in the corresponding filter. This luminosity was used to calibrate the final 1D spectra. During this process, we also identified a prominent host galaxy in J1625+4309. The details of the 2D image decomposition are presented in Section~\ref{sec:host}.

The final 1D spectra preserve the relative flux calibration between components while minimizing cross-contamination, enabling accurate line profile analysis.

% \begin{figure*}[htbp!]
%      \centering
%      \gridline{\fig{target1.png}{0.9\textwidth}{(a) J1625+4309}}
%      \gridline{\fig{target2.png}{0.9\textwidth}{(b) J0229$-$0514}}
%      \caption{
%      The HST observation data for two confirmed targets: J1625+4309 in panel (a) and J0229$-$0514 in panel (b). In each panel, we show the direct imaging and spectral data as follows: {\it Left:} Direct image in the F140W filter obtained with HST/WFC3, showing the spatial configuration of the two components. {\it Upper right:} Two-dimensional slitless spectrum extracted with HSTaXe, oriented such that the dispersion axis runs horizontally, consistent with the direct image orientation, and the two boxes in the image indicate the extraction range of the one-dimensional spectrum. {\it Lower right:} Final extracted one-dimensional spectra of the two components. The four quasars in two dual systems are labeled as ID1a, ID1b, ID2a, and ID2b.}
%      \label{fig:data_reducted}
% \end{figure*}

\section{Dual quasar confirmation} \label{sec:confirm}

Spectroscopic analysis confirms two dual quasar systems: J1625+4309 ($z=1.647$, separation 0\farcs55/4.7 kpc), with both components (named ID1a and ID1b hereafter) displaying prominent and spatially distinct H$\beta$ emission, and J0229$-$0514 ($z = 3.174$, separation 0\farcs42/3.2 kpc), with both components (named ID2a and ID2b hereafter) exhibiting broad Mg\,\textsc{ii} lines. Given to their redshifts (see Section~\ref{sec:sample}), both targets were observed by F140W/G141.
The remaining nine candidates do not show spectroscopic evidence for two active nuclei, and thus are not classified as dual quasars. Their data can be found in the Appendix \ref{app:non_detect}.%This final confirmed rate (2/59,025) suggests dual quasars with sub-6~kpc separations at cosmic noon are exceptionally rare, occurring in approximately $10^{-4}$ of the luminous quasar population.

Figure~\ref{fig:data_reducted} presents the HST observational data of the two systems, with subfigure (a) corresponding to J1625+4309 and subfigure (b) corresponding to J0229$-$0514. 
For J1625+4309, the one-dimensional spectrum of ID1a and ID1b reveals prominent emission lines of H$\beta$ and [O\,\textsc{iii}], which are characteristic features of quasars. The clear separation and distinct profiles of these emission lines in both components provide strong evidence that J1625+4309 is a dual-quasar system. The direct HST imaging reveals a clear spatial separation between the two quasars, with no evidence for a foreground lensing galaxy or extended arcs typically associated with strong gravitational lensing. Additionally, the emission lines in the spectra of the two components are distinct in both line shape and velocity offset. These features demonstrate that this is a genuine physical pair, rather than the result of lensing or chance alignment.

J0229$-$0514 is similarly confirmed as a dual quasar system: both components are spatially resolved in the direct imaging, with no indication of a lensing galaxy or arc-like features. The spectra show broad Mg\,\textsc{ii} lines in both nuclei, but with a clear velocity offset ($\sim$1000~km/s), unambiguously excluding a lensing interpretation and demonstrating the system’s physical dual nature.

\begin{figure*}[htbp!]
     \centering
     \includegraphics[width=0.9\textwidth]{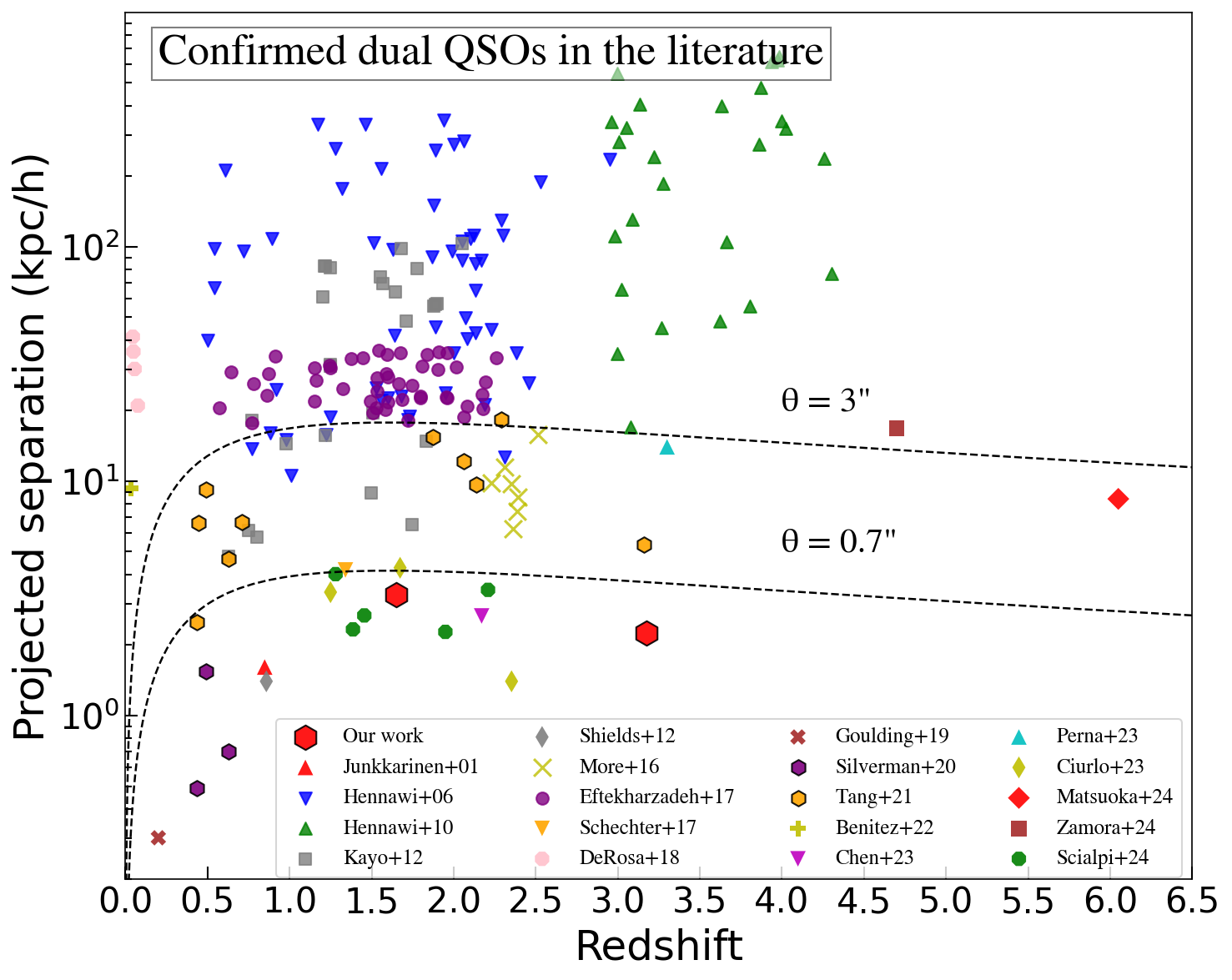}
     \caption{Redshift versus projected separation for confirmed dual quasars in the literature. Our newly identified systems (red hexagons) populate the sparsely sampled parameter space at $1<z<3$ with separations $<5$ kpc, extending the sample below the $\theta=0.\!\!^{\prime\prime}7$ angular separation limit (lower dashed curve). Many early certification results are concentrated above $\theta \sim 3^{\prime\prime}$ (upper dashed curve), reflecting a limitation based on credibility selection. Symbol types denote different discovery programs: \cite{Junkkarinen_2001, hennawi_2006, hennawi_2010, kayo_2012, shields_2012, more_2016, Schechter_2017, eftekharzadeh_2017, derosa_2018, goulding_2019, Silverman_2020, tang_2021, Mannucci_2022, benitez_2022, Ciurlo_2023, Chen_2023, Perna_2023, matsuoka_2024, Scialpi_2024, zamora_2024}, and this work. J0229$-$0514 at $z=3.174$ (lower right hexagon) represents the highest-redshift confirmed sub-5 kpc dual quasar to date.
     }
     \label{fig:QSOs_confirmed}
\end{figure*}

Figure~\ref{fig:QSOs_confirmed} presents a compilation of all confirmed dual quasars reported in the literature to date, showing their redshifts against projected physical separations. Our HST observations expand the parameter space of confirmed dual quasars at $1 < z < 3$ with separations $<0\farcs7$ (red hexagons in the figure). Notably, at this separation range, J0229$-$0514 at $z=3.174$ remains the only confirmed dual quasar at $z>3$ so far.

We note that several other studies have reported candidate close-separation AGN pairs \citep{Glikman_2023, Barrows_2012} that were not included in Figure 3 as they lack spectroscopic confirmation of dual AGN activity. Similarly, \cite{Perna_2023} identified several quasar pairs, but most lack confirmation of dual AGN nature. By contrast, our work provides robust confirmation through matched broad-line emission in both components.

\subsection{Direct Image Decomposition} \label{sec:host}
We performed two-dimensional quasar–host galaxy decomposition for both systems using \texttt{Galight} on the HST direct imaging data to search for host galaxy emission for each component. This modeling also quantifies the fractional contribution of the quasar point sources to the total system luminosity, providing a robust flux calibration for the extracted one-dimensional spectra. The corresponding correction factors were applied to scale the spectral data appropriately.

\begin{figure*}[htbp!]
    \centering

\begin{minipage}{0.9\textwidth}
\centering
\includegraphics[width=\textwidth]{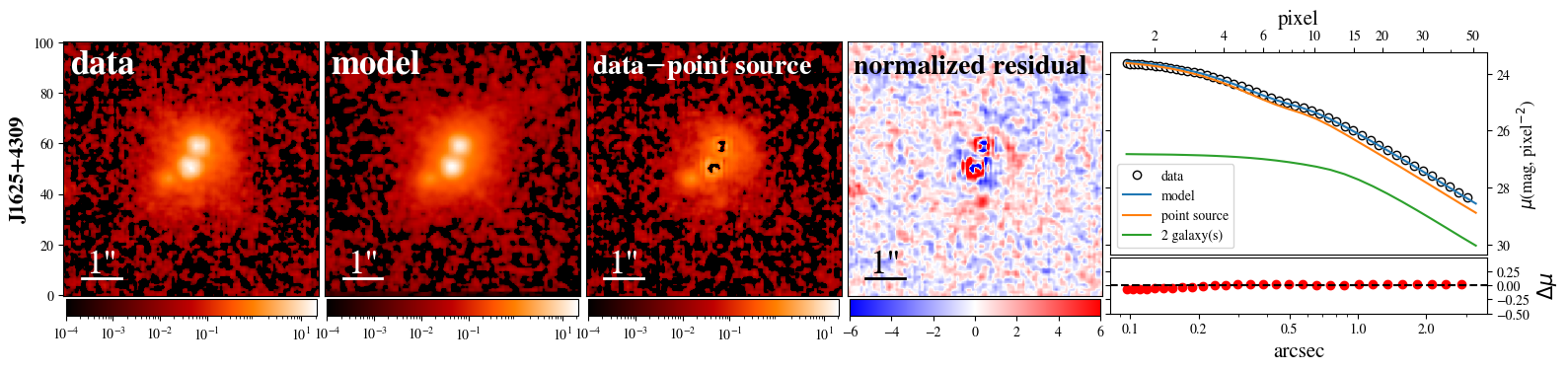}
\\(a) J1625+4309
\end{minipage}

%\vspace{0.3cm}

\begin{minipage}{0.9\textwidth}
\centering
\includegraphics[width=\textwidth]{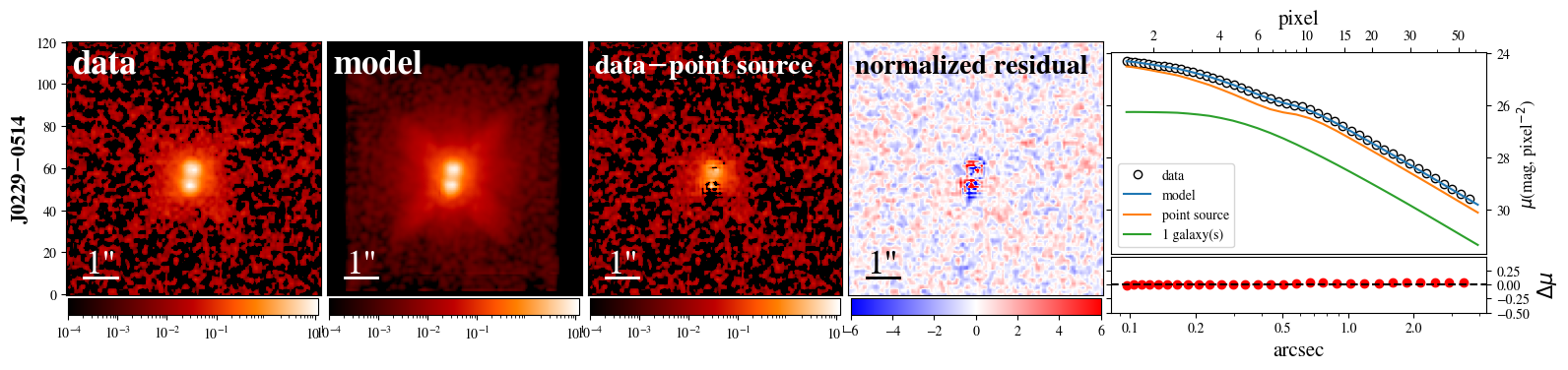}
\\(b) J0229$-$0514
\end{minipage}

\caption{Multi-component decomposition of J1625+4309 and J0229-0514 in the F140W band. For each subgraph: \textit{Left to right}: (i) Original HST/WFC3 direct image; (ii) Best-fit quasar point-source model from \texttt{Galight}; (iii) Residual map after quasar subtraction; (iv) Normalized residuals ($\mathrm{Data - Model}$)/$\sigma$; (v) Demonstrates the radial surface brightness curve from the center of the image of each cut in 1D. The scale is already marked in the figure.}
\label{fig:galight_fit}
\end{figure*}

For both systems, we performed decomposition using a model consisting of two point-source components (for the dual quasars) and one primary Sérsic profile (for the host galaxy). The PSF models used in the decomposition were constructed from isolated stars within the field of view, ensuring that they accurately represents the observational conditions.  For J1625+4309, we added a secondary Sérsic component to the model to represent the prominent nearby target visible in the lower-left region of the host galaxy.
The results of the decomposition are presented in Table \ref{tab:properties}. 

The results of the decomposition for the two systems are displayed in Figure~\ref{fig:galight_fit}, which shows the original direct images, the best-fit models, and the residual maps. For J1625+4309, the system exhibits a complex and disturbed morphology consistent with an ongoing major merger. The fit gives a separation in the F140W band of 0.57\arcsec. The effective radius of the host galaxy is 0.56\arcsec. %\todo{add ~kpc value}. 
The nearby target visible in the lower-left region of the system may correspond to a major tidal feature, an infalling companion, or enhanced star formation triggered by the interaction. These results support the scenario of dual AGN residing within interacting galaxies. The clear separation of light from each nucleus also enables a robust assessment of the host galaxy structure and the AGN–host flux ratio, which is critical for calibrating the quasar luminosity to measure black hole masses.

In contrast, for J0229-0514, with a 0.50\arcsec~separation of the dual quasars, the host galaxy is more compact ($R_e$=0.18\arcsec/1.4 kpc)%\todo{(add kpc size)}
, indicating that this system could be at a more advanced merger stage, with the host galaxies already coalesced or the most prominent tidal features having faded.

To assess the robustness of the host galaxy measurement, particularly for J0229$-$0514 which has a compact host galaxy, we additionally tested different PSF models extracted from the same observing field of view. Our inferred results consistently recover the host galaxy component. For instance, in the case of J0229$-$0514, the derived fluxes agree within 0.3 mag, while the effective radii remain stable within $0.6$ kpc.

The JWST/NIRSpec Integral Field Unit (IFU) will be extremely valuable for understanding the system by providing spatially resolved spectroscopy across the object in a single observation. This capability allows for simultaneous imaging and spectral analysis, enabling detailed mapping of kinematics, emission lines, and the physical conditions within different regions of the system.

\subsection{Black Hole Mass Estimates} \label{sec:estimates}

\begin{figure*}[htbp!]
\centering

\begin{minipage}{0.9\textwidth}
\centering
\includegraphics[width=\textwidth]{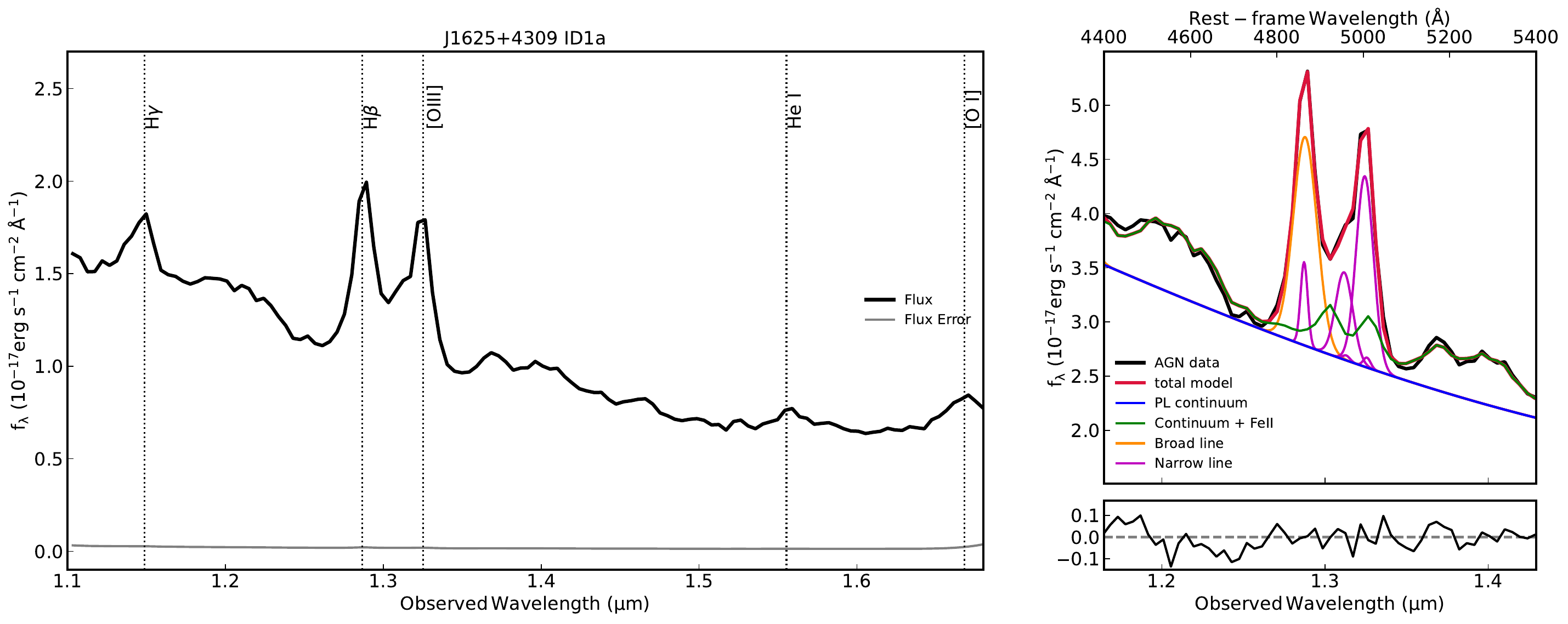}
\\(a)
\end{minipage}

%\vspace{0.3cm}

\begin{minipage}{0.9\textwidth}
\centering
\includegraphics[width=\textwidth]{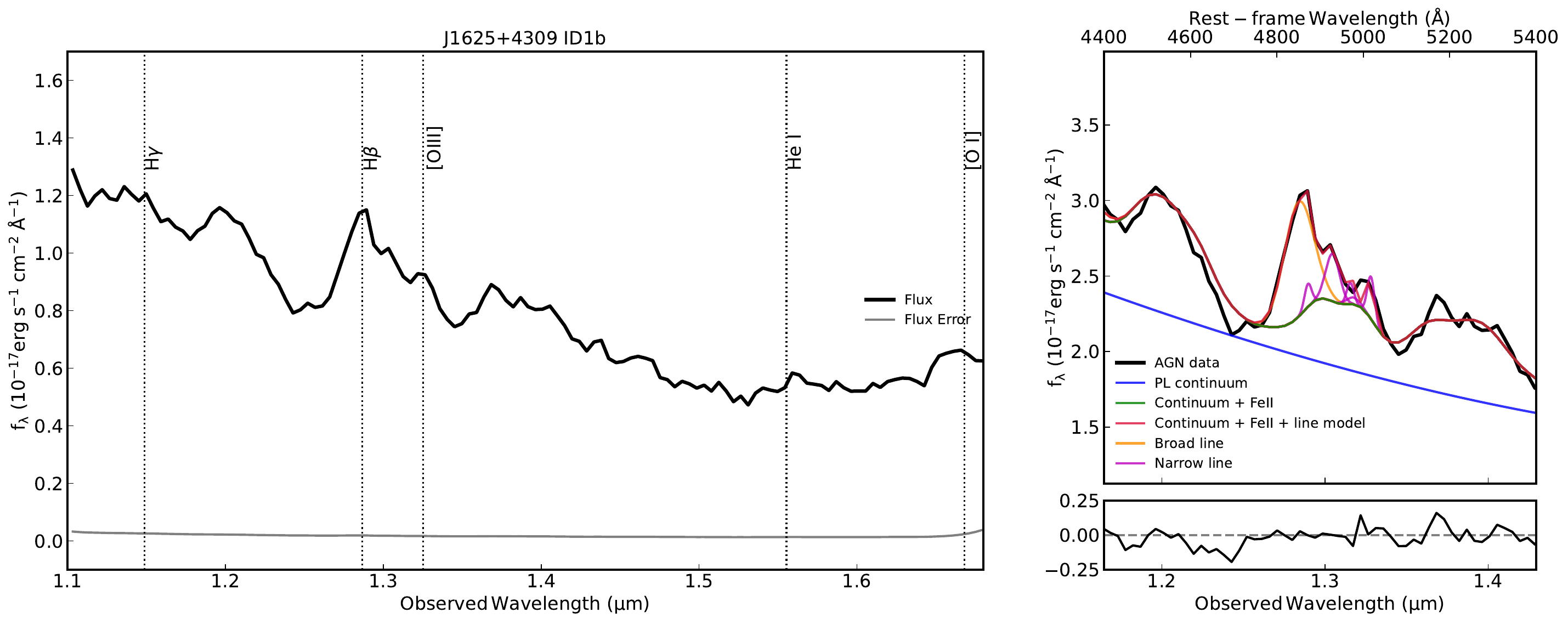}
\\(b)
\end{minipage}

\caption{Spectral fitting results for J1625+4309. For each panel: \textit{Left}: Calibrated 1D spectrum with identified emission lines within the observed wavelength range, with emission lines position indicated. \textit{Upper right}: The results of the spectral fitting focused on the H$\beta$+[O\,\textsc{iii}] region for the dual quasar systems. The multi-component decomposition including: power-law continuum (blue), Fe~II template (where detected, green), broad emission lines (orange), and narrow lines (magenta). \textit{Lower right}: Residuals of the spectral fit.}
\label{fig:qso_fit1}
\end{figure*}

\begin{figure*}[htbp!]
\centering

\begin{minipage}{0.9\textwidth}
\centering
\includegraphics[width=\textwidth]{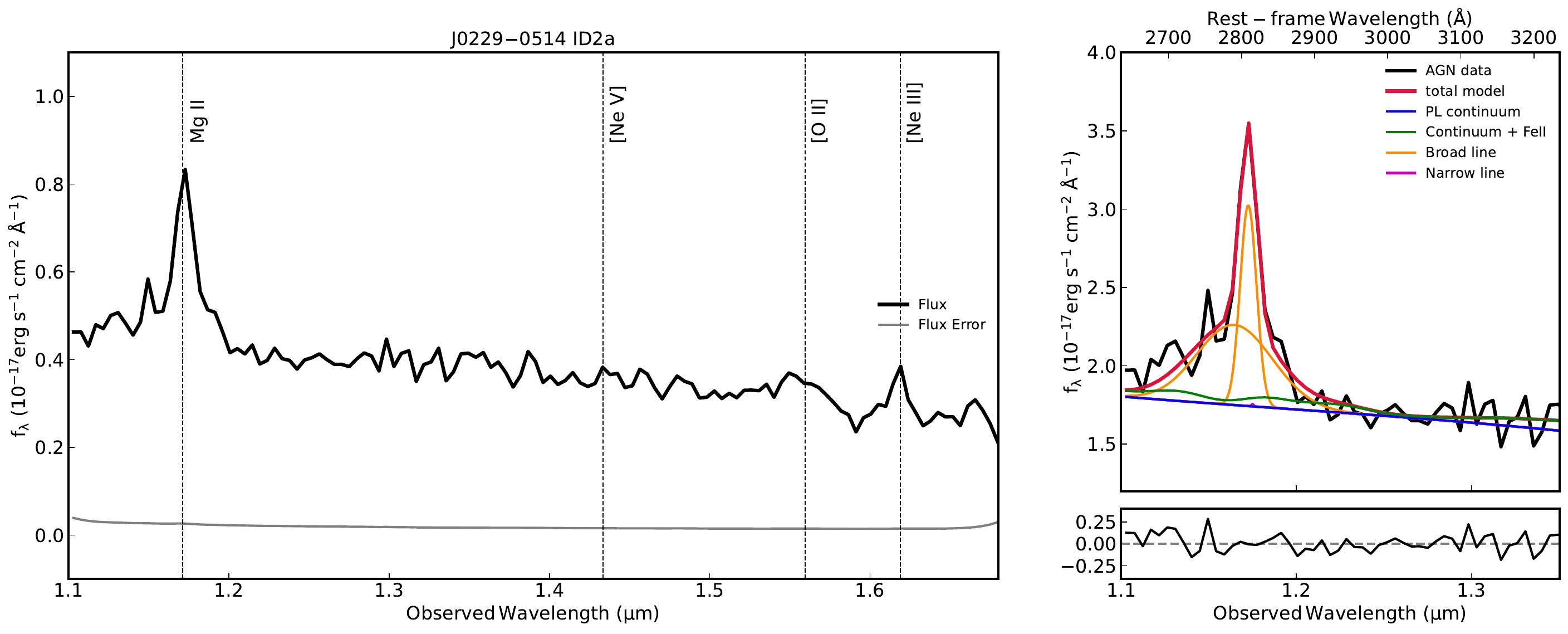}
\\(a)
\end{minipage}

%\vspace{0.3cm}

\begin{minipage}{0.9\textwidth}
\centering
\includegraphics[width=\textwidth]{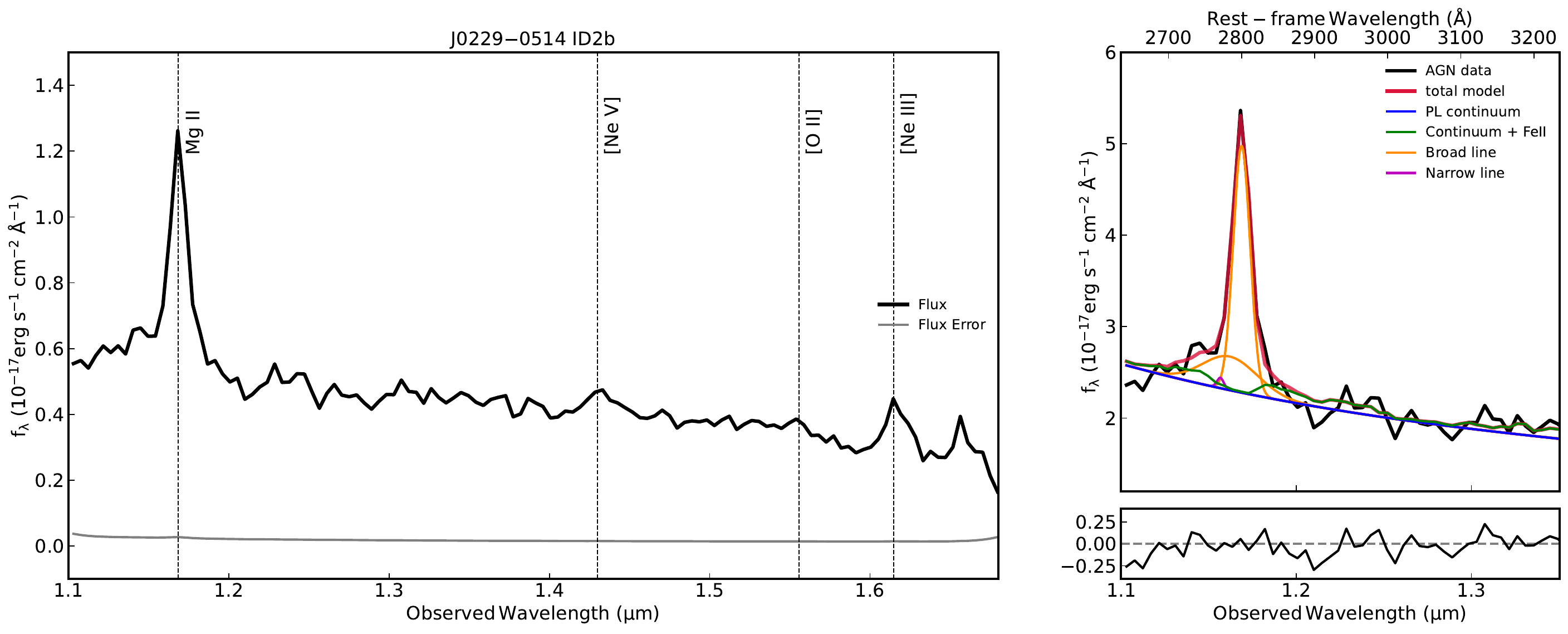}
\\(b)
\end{minipage}

\caption{Spectral fitting results for J0229$-$0514, with a same panel arrangement as \ref{fig:qso_fit1} but spectral fitting focus on Mg\,\textsc{ii}.}
\label{fig:qso_fit2}
\end{figure*}

The derived one-dimensional spectra extracted from HST slitless data by our method (i.e., 3-pixel region described in Section~\ref{sec:reduc}) do not fully capture the total flux from each quasar, as slitless extraction can omit emission outside the chosen aperture. To obtain accurate black hole mass estimates, we use our F140W direct imaging and its two-dimensional decomposition to quantify the total quasar luminosity that should enter the spectrograph for each nucleus. By scaling the 1D spectra with these imaging-derived correction factors, we recover the intrinsic flux for each component. 

To obtain precise emission line information from the active galactic nucleus (AGN), we performed spectral fitting on the candidate objects using the Python package \texttt{PyQSOFit} \citep {guo_2018,shen_2019,ren_2024}. This software is specifically designed for AGN spectral fitting and can simultaneously handle both continuum and emission line components. 

For J1625+4309, we include major emission lines such as H$\beta$, [O\,\textsc{iii}]$\lambda$5007 and [O\,\textsc{iii}]$\lambda$4959 during the fitting process, while we focus on the Mg\,\textsc{ii} emission line in J0229$-$0514, along with continuum components such as power-law continuum and iron emission templates (see Figure~\ref{fig:qso_fit1}). More specifically, the emission line fitting adopted a multi-component approach: broad and narrow components for H$\beta$/Mg\,\textsc{ii}, and a core+wing configuration for [O\,\textsc{iii}] to capture its non-Gaussian asymmetry.

We note that the host galaxy component is not well constrained in any of the spectral fits. This limitation is primarily due to the low signal-to-noise ratio in the continuum and incomplete wavelength coverage of key stellar absorption features. Nevertheless, our F140W direct imaging decomposition shows that, even for J1625+4309, which has a prominent extended host, the AGN emission overwhelmingly dominates the total system light (see Figure~\ref{fig:galight_fit}). To ensure accurate quasar continuum measurements, we have re-calibrated the AGN luminosity in the spectra using the AGN-only component derived from direct imaging to ensure an unbiased estimation of the continuum luminosity for black hole mass determination.

\begin{table}
\caption{Imaging and Spectroscopic Measurements}
\label{tab:properties}
% \centering

% \begin{center}

% \hspace*{-0.5in}
% \begin{adjustbox}{width=0.55\textwidth}
% \begin{tabular}{p{0.35\linewidth}cc}

\resizebox{1.05\columnwidth}{!}{
\centering
\hspace{-0.6in}
\begin{tabular}{lcc}
\toprule
\multirow{2}{*}{\textbf{Property}} & \textbf{J1625+4309} & \textbf{J0229$-$0514} \\
& \textbf{(ID1a/b)} & \textbf{(ID2a/b)} \\
\midrule

RA, DEC (deg) & 246.258255, 43.1587791 & 37.2751989, -5.2413844 \\
\midrule

\textbf{Imaging} & & \\
mag$_{\mathrm{F140W}}$ & 19.7, 20.0 & 20.6, 20.7 \\
Separation & 0.57$''$ (4.8 kpc) & 0.50$''$ (3.8 kpc) \\
Host Sérsic & 0.7 & 3.7 \\
Host $R_e$ & 0.56$''$ (4.7 kpc) & 0.18$''$ (1.4 kpc) \\
Host mag$_{\mathrm{F140W}}$ & 20.4 & 21.2 \\
\midrule

\textbf{Spectroscopic} & & \\
FWHM$_{\mathrm{H}\beta}$ & $3.4\pm0.7$, $4.0\pm1.0$ & -- \\
FWHM$_{[\mathrm{O}\,\textsc{iii}]}$ & $1.7\pm0.2$, $1.6\pm0.4$ & -- \\
FWHM$_{\mathrm{Mg\,\textsc{ii}}}$ & -- & $2.9\pm0.4$, $2.2\pm0.2$ \\
$\log\mathcal{L}$ (erg s$^{-1}$) & 45.29, 45.15 & 45.65, 45.73 \\
$\log(M_{\mathrm{BH}}/M_\odot)$ & 8.6$\pm$0.2, 8.7$\pm$0.2 & 8.3$\pm$0.1, 8.1$\pm$0.1 \\
$v_{\mathrm{Los}}$ (km/s) & $(0.7\pm0.1)\times10^{3}$ & $(1.0\pm0.2)\times10^{3}$ \\
\bottomrule
\end{tabular}

}
    
% \end{center}
% \end{adjustbox}

\tablecomments{The inferred properties for the dual quasars. FWHM in $10^{3}$ km/s. For H$\beta$ and Mg\,\textsc{ii}, the $\mathcal{L}$ are presented for $\mathcal{L}_{\rm 5100}$ and ${\mathcal{L}_{\rm 3000}}$, respectively.}
\end{table}

Table~\ref{tab:properties} presents key measurements from our spectroscopic analysis. The FWHM values for both pairs range from $(2.2\pm0.2)\times10^{3}$~km~s$^{-1}$ to $(4.0\pm1.0)\times10^{3}$~km~s$^{-1}$, consistent with Type-1 AGN broad-line regions, which provides further confirmation that both systems are genuine physical dual quasars.

The mass of the central supermassive black hole ($\mathcal{M}_{\mathrm{BH}}$) can be 
estimated using the virial theorem applied to broad emission lines from the active galactic 
nucleus. We employ the following empirical relation:

\begin{eqnarray} 
\log\left(\frac{\mathcal{M}_{\mathrm{BH}}}{M_{\odot}}\right) & 
= a+b\log\left(\frac{\mathcal{L}_{\lambda_{\mathrm{line}}}}{10^{44}\mathrm{erg~s}^{-1}}\right) 
\nonumber\\ & +2\log\left(\frac{\mathrm{FWHM}(\mathrm{line})}{1000\mathrm{~km~s}^{-1}}\right),
\end{eqnarray}
where $\lambda_{\mathrm{line}}$ is the reference wavelength of the local continuum luminosities for different emission lines (H$\beta$ or Mg\,\textsc{ii}), and FWHM(line) is its full width at half maximum. The coefficients $(a, b)$ are adopted from \citet{Ding_2020} as $(a, b) = (6.910, 0.50)$ for H$\beta$ and $(a, b) = (6.623, 0.47)$ for Mg\,\textsc{ii}. %\todo{confirm these values.} 
Notably, the FWHM values reported in the work have been corrected for the instrumental broadening inherent in the HST G141 grism (R $\approx 130$, corresponding to 2300 km s$^{-1}$). The final calibrated BH mass estimates span $10^{8.1}-10^{8.7}~M_{\odot}$, with the J1625+4309 system hosting more massive black holes than J0229$-$0514.

The line-of-sight velocity offsets ($\Delta v_{\rm LOS}$) between the dual quasar components were measured from the centroid shifts of their broad emission lines. We find velocity separations of $(0.7\pm0.1)\times10^{3}$ km s$^{-1}$ for J1625+4309 and $(1.0\pm0.2)\times10^{3}$ km s$^{-1}$ for J0229$-$0514, consistent with their identification as physically bound systems (velocity differences $<$2000 km $s^{-1}$) rather than chance projections or lensing duplicates.

\section{Summary} \label{sec:summary}

We present a systematic HST/WFC3 observation for sub-arcsecond dual quasars at $z > 1$, combining high-resolution imaging (0\farcs06 resolution) and slitless grism spectroscopy to overcome previous observational limitations. From an initial sample of 59,025 SDSS quasars in the HSC-SSP footprint, we identified 386 candidate pairs through PSF-fitting analysis and selected 11 high-priority targets with separations $<0\farcs7$ ($<6$ kpc) for HST follow-up. Our analysis confirms two robust dual quasar systems:

1. SDSS J1625+4309 at $z=1.647$ with separation 0\farcs55 (4.7 kpc), exhibiting matching broad H$\beta$ emission in both components (FWHM = $(3.4\pm0.7)\times10^{3}$ km s$^{-1}$ and $(4.0\pm1.0)\times10^{3}$ km s$^{-1}$), $\mathcal{M}_{\rm BH} = 10^{8.6} M_\odot$ and $10^{8.7} M_\odot$, and a line-of-sight velocity offset of $(0.7\pm0.1)\times10^{3}$ km s$^{-1}$.

2. SDSS J0229$-$0514 at $z=3.174$ with separation 0\farcs42 (3.2 kpc), showing broad Mg\,\textsc{ii} emission in both nuclei (FWHM = $(2.9\pm0.4)\times10^{3}$ km s$^{-1}$ and $(2.2\pm0.2)\times10^{3}$ km s$^{-1}$), $\mathcal{M}_{\rm BH} = 10^{8.3} M_\odot$ and $10^{8.1} M_\odot$, and velocity offset $(1.0\pm0.2)\times10^{3}$ km s$^{-1}$.

The host galaxy analysis (i.e., Figure~\ref{fig:galight_fit}) reveals distinct merger stages: J1625+4309 shows an extended ($R_e = 4.7$ kpc), disturbed morphology indicative of an ongoing major merger, while J0229$-$0514 exhibits a compact host ($R_e = 1.4$ kpc) suggestive of advanced coalescence. 

% These discoveries represent the first confirmed dual quasars with sub-5 kpc separations at $z>1.5$, filling a critical gap in our understanding of SMBH binary evolution.

These discoveries add new examples of confirmed dual quasars with sub-5 kpc separations at $z>1.5$, helping to fill the critical gap in our understanding of SMBH binary evolution.

Our novel methodology, combining HST's angular resolution with spectral confirmation of both components, provides a powerful approach for future dual AGN searches. These systems represent prime targets for JWST/NIRSpec IFU follow-up to map merger-induced gas flows.

\begin{acknowledgments}

Based in part on observations made with the NASA/ESA
Hubble Space Telescope, obtained at the Space Telescope
Science Institute, which is operated by the Association of
Universities for Research in Astronomy, Inc., under NASA
contract NAS 5-26555. These observations are associated with
program No. 17143. Support for this work was provided by
NASA through grant No. HST-GO-17143 from the Space
Telescope Science Institute, which is operated by AURA, Inc.,
under NASA contract NAS 5-26555.
All of the data presented in this article were obtained from the Mikulski Archive for Space Telescopes (MAST) at the Space Telescope Science Institute. The specific observations analyzed can be accessed via \dataset[doi:10.17909/36dv-p141]{https://doi.org/10.17909/36dv-p141}.

We thank Luis C. Ho and Jianghua Wu for helpful suggestions that improved this paper. 

X.D. and K.L. acknowledge the National Key R\&D Program of China (No. 2024YFC2207400). K.L. acknowledges the National Natural Science Foundation of China (NSFC) No. 12222302.
K.K. acknowledges support from JSPS KAKENHI Grant Numbers JP22H04939, JP23K20035, and JP24H00004.

\end{acknowledgments}

\appendix

\section{Overview of non-detected dual-QSO}
\label{app:non_detect}
\renewcommand{\thefigure}{A.\arabic{figure}} 
\setcounter{figure}{0}

\begin{figure}[h]
\centering

\begin{minipage}{0.48\textwidth}
    \centering
    \includegraphics[width=\linewidth]{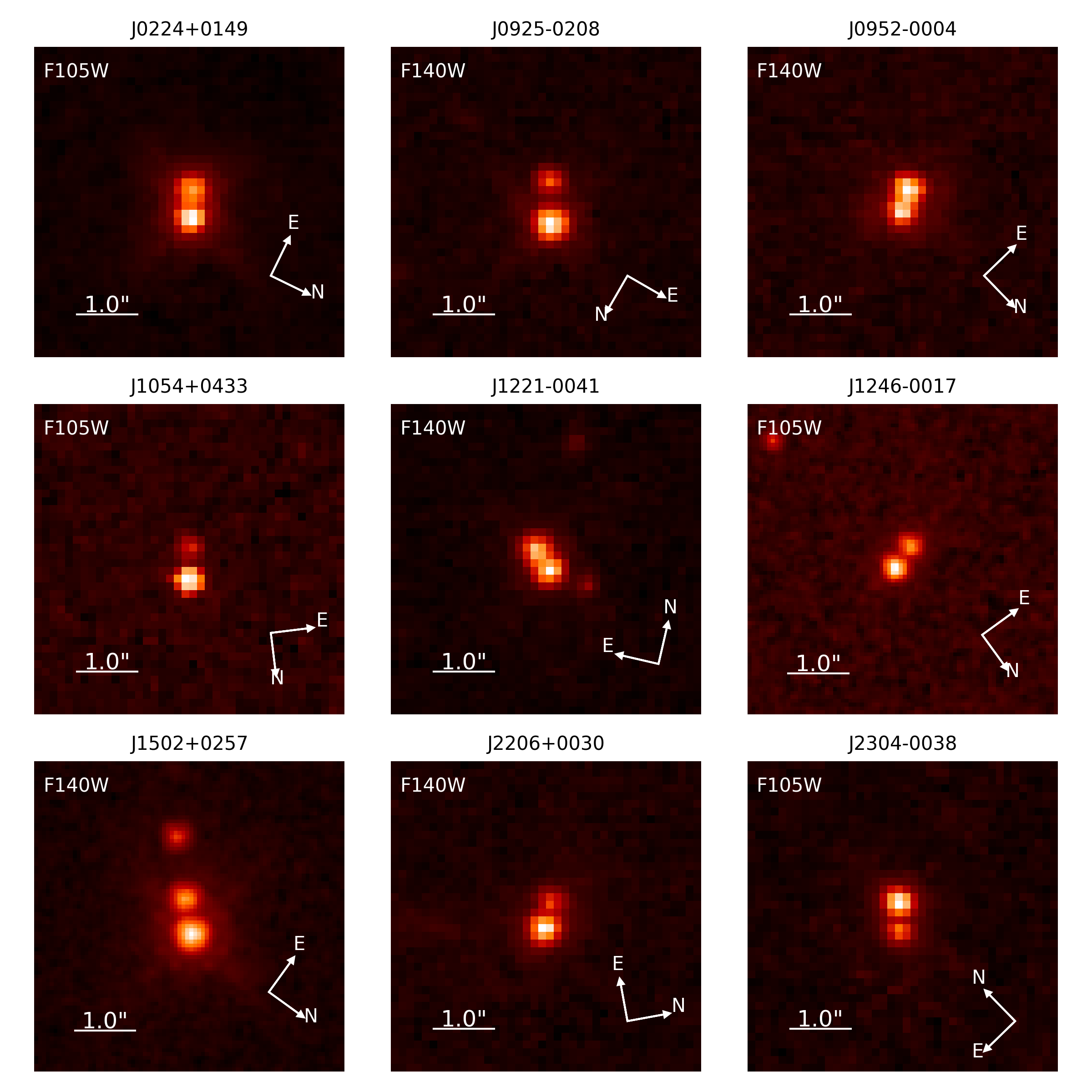}
    \par\vspace{0.5em} 
    \textbf{(a)} imaging data of non-detected dual-QSOs
    \label{fig:non_dual_QSO_image}
\end{minipage}
\hfill
\begin{minipage}{0.48\textwidth}
    \centering
    \includegraphics[width=\linewidth]{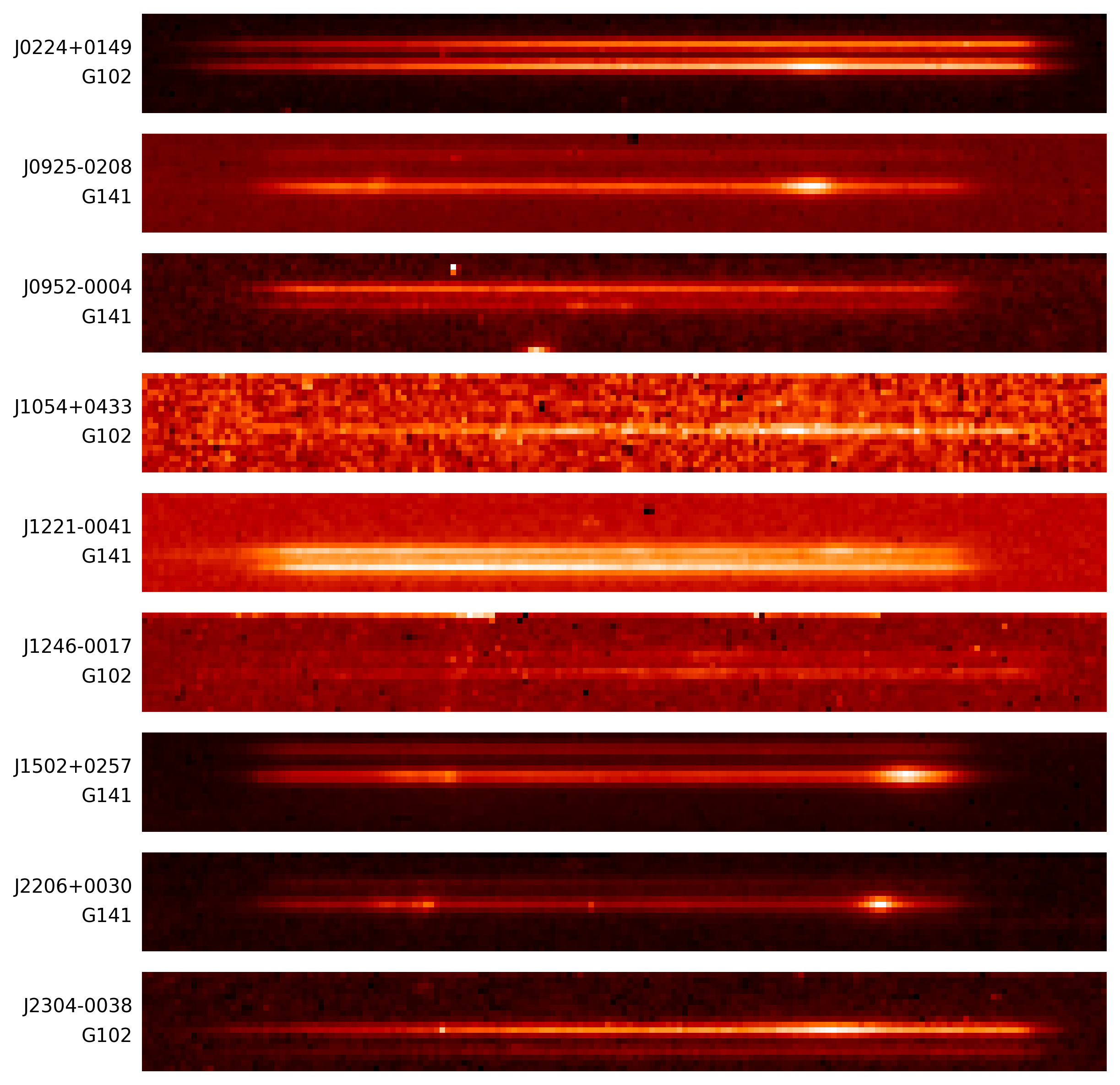}
    \par\vspace{0.5em}
    \textbf{(b)} spectra data of non-detected dual-QSOs
    \label{fig:non_dual_QSO_spec}
\end{minipage}

\vspace{0.5cm}

\begin{minipage}{0.9\textwidth}
    \centering
    \includegraphics[width=\linewidth]{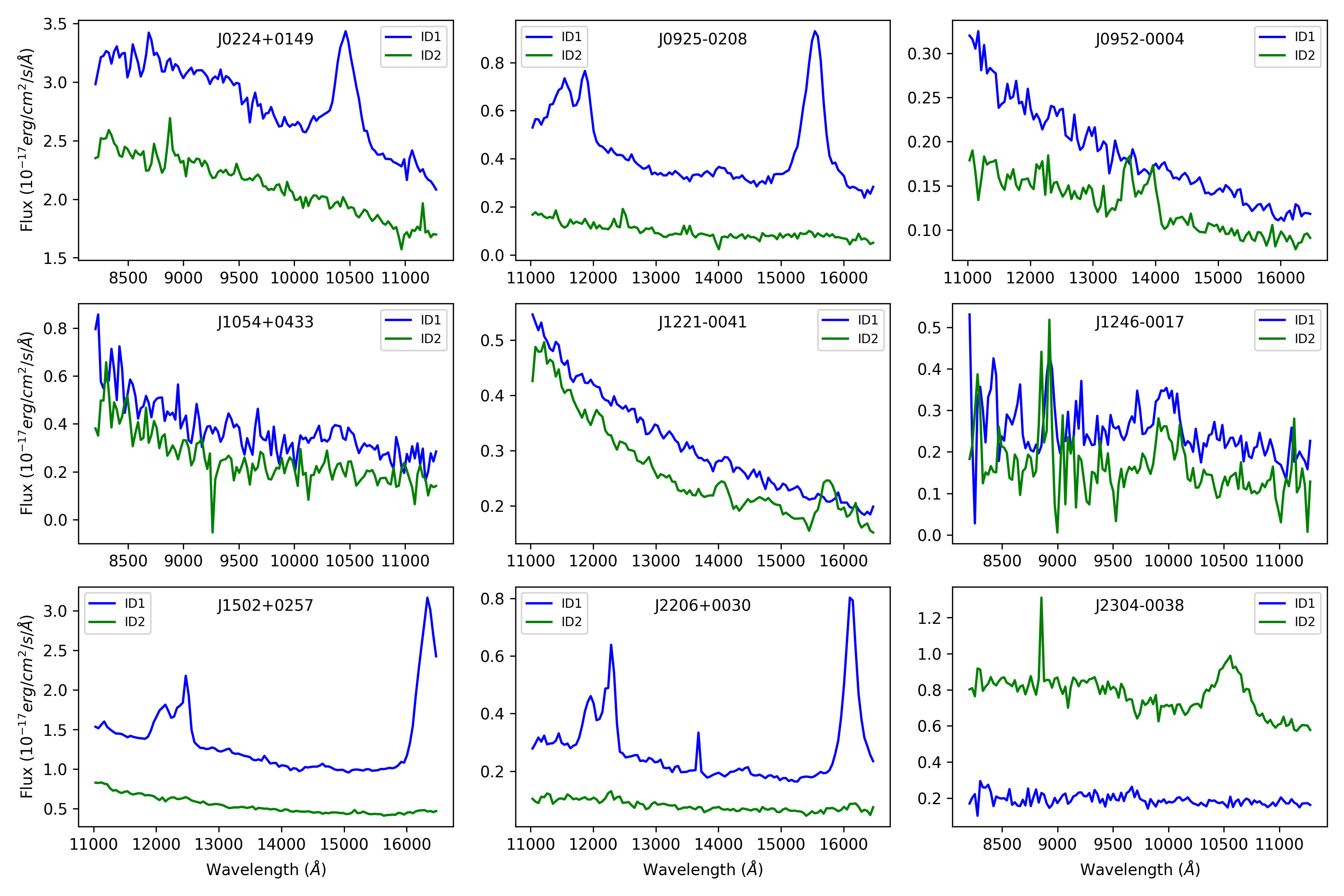}
    \par\vspace{0.5em}
    \textbf{(c)} 1D spectra data of non-detected dual-QSOs
    \label{fig:non_dual_QSO_spec1D}
\end{minipage}

\caption{These present data from nine other sources that have not been detected as dual quasars, which will be discussed in detail regarding their properties in future article. These are: (a) direct imaging data, (b) two-dimensional spectral data, (c) one-dimensional spectral data.}
\label{fig:non_detect}
\end{figure}

%\bibliography{sample631}{}

\bibliographystyle{aasjournal}

\end{document}